\documentclass[journal,twoside]{IEEEtran}
\usepackage{amssymb}
\usepackage{multirow}
\usepackage{indentfirst}
\usepackage{inputenc}
\usepackage{enumerate} 
\usepackage{textcomp}
\usepackage{listings}
\usepackage{array}
\usepackage[switch,mathlines]{lineno}
\usepackage{soul}
\usepackage{xfrac}
\usepackage{graphicx}
\usepackage{cite}
\usepackage[cmex10]{amsmath}
\usepackage{algorithm}
\usepackage{stackengine}

\usepackage{algpseudocode}
\usepackage{fixltx2e}
\usepackage{CJK}
\usepackage[cmex10]{amsmath}
\usepackage{bm}
\usepackage{color}
\usepackage{amsmath,amsthm,amssymb,amsfonts}
\usepackage{flushend}
\usepackage{float}

\usepackage{arydshln} 
\usepackage{hyperref}
\hypersetup{
     colorlinks   = true,
     linkcolor    = blue,
     citecolor    = red
}
\makeatletter
\newcommand*{\transpose}{%
  {\mathpalette\@transpose{}}%
}
\newcommand*{\@transpose}[2]{%
  \raisebox{\depth}{$\m@th#1\intercal$}%
}
\makeatother

\usepackage{soul}
\usepackage{tikz}
\usepackage{booktabs}
\usepackage{tabularx}

\DeclareMathOperator{\Cov}{Cov}

\usepackage[caption=false,font=footnotesize]{subfig}
\usepackage{fixltx2e}
\ifCLASSINFOpdf
\else
\fi

\begin{document}

\bstctlcite{IEEEexample:BSTcontrol}

\title{ Probabilistic Load-Margin Assessment using  Vine Copula and Gaussian Process Emulation}



\author{
{Yijun~Xu,~\IEEEmembership{Member, IEEE}, 
Kiran~Karra,
Lamine~Mili,~\IEEEmembership{Life Fellow, IEEE},
Mert~Korkali,~\IEEEmembership{Senior Member, IEEE},
Xiao~Chen,
Zhixiong~Hu}
       \thanks{Y. Xu and L. Mili are with the Bradley Department of Electrical and Computer Engineering, Virginia Tech, Northern Virginia Center, Falls Church, VA 22043 USA (e-mail:\{yijunxu,lmili\}@vt.edu).}
 \thanks{K. Karra is with the Applied Physics Lab at Johns Hopkins University  
Baltimore, MD 21218 USA (e-mail: kkarranc@vt.edu).}
 \thanks{M.~Korkali is with the Computational Engineering Division, Lawrence Livermore National Laboratory, Livermore, CA 94550 USA (e-mail: korkali1@llnl.gov).}
 \thanks{X.~Chen is with the Center for Applied Scientific Computing, Lawrence Livermore National Laboratory, Livermore, CA 94550 USA (e-mail: chen73@llnl.gov)}
  \thanks{Z. Hu is with the Department of Statistics, University of California-Santa Cruz, 
Santa Cruz, CA 95064 USA (e-mail: zhu95@ucsc.edu).}
\thanks{This work was supported, in part, by the U.S. National Science Foundation under  Grant 1917308 and by the United States Department of Energy Office of Electricity Advanced Grid Modeling Program, and performed under the auspices of the U.S. Department of Energy by Lawrence Livermore National Laboratory under Contract DE-AC52-07NA27344. Document released as LLNL-PROC-795980.}

}

\markboth{IEEE PES General Meeting, August 2-6, 2020 Montreal, Canada}%
{Xu \MakeLowercase{\textit{\textit{et al.}}}: Probabilistic Load-Margin Assessment with  Vine Copula and Gaussian Process Emulation}
\maketitle
\begin{abstract}
The increasing penetration of renewable energy along with the variations of the loads bring large uncertainties in the power system states that are threatening the security of power system planning and operation.  Facing these challenges, this paper proposes a cost-effective, nonparametric method to quantify the impact of uncertain power injections  on the load margins. First, we propose to generate system uncertain inputs via a novel vine copula due to its capability in simulating complex multivariate highly dependent model inputs. Furthermore, to reduce the prohibitive computational time required in the traditional Monte-Carlo method, we propose to use a nonparametric, Gaussian-process-emulator-based reduced-order model to replace the original complicated continuation power-flow model.  This emulator allows us to execute the time-consuming continuation power-flow solver at the sampled values with a negligible computational cost. The simulations conducted on the IEEE 57-bus system, to which correlated renewable generation are attached, reveal the excellent performance of the proposed method.

\end{abstract}
\IEEEpeerreviewmaketitle
\vspace{-0.3cm}
\section{Introduction}
\IEEEPARstart{P}OWER  systems exhibit stochastic dynamics, in particular due to the continuous variations of the loads  and the intermittency of the renewable generation, among other causes. To address this problem, research activities have focused on uncertainty assessment in power system planning, operation, and control. Examples include a probabilistic power flow \cite{fan2012probabilistic,xu2019risk}, an uncertainty quantification for power system dynamic simulation \cite{xu2018propagating}, a stochastic economic dispatch\cite{safta2016efficient, hu2019uncertainty}, and a probabilistic load-margin formulation\cite{haesen2009probabilistic,xu2017power}, among others.  The latter formulation is critical to ensuring the voltage stability of modern power systems with increasing penetration of renewables and, therefore, is chosen as the scope of this paper.

Traditionally, Monte-Carlo (MC) simulations have been utilized to address this problem \cite{da2000voltage} . However, the computing time of a single continuation power flow (CPF) case is much longer than that of a simple power-flow case. Therefore, the straightforward MC method based on the evaluations of tens of thousands samples in the CPF model
will be prohibitively time-consuming. This is even true for relatively small systems. Although some analytical methods have been proposed to reduce the computational burden,  there exists no very accurate closed-form solution today due to the nonlinearity of the CPF model~\cite{haesen2009probabilistic}.

To overcome the abovementioned shortcomings, this paper proposes, for the first time, to utilize a method based on a Gaussian process emulator (GPE) to solve the probabilistic CPF problem. Known as a Bayesian-learning-based method for a nonlinear regression problem in statistics~\cite{kennedy2001bayesian}, the GPE can serve as a nonparametric, reduced-order model representation for the nonlinear CPF model. This emulator allow us to evaluate the time-consuming CPF solver at the sampled values with a negligible computational cost. 
Furthermore, to simulate the high-dimensional dependent samples that represent the uncertainties from the loads and renewables, we propose to adopt a novel 
vine-copula technique \cite{konstantelos2018using}, which is capable of modeling the high-dimensional, dependent multivariate with a variety of bivariate copulae  such as Frank and Gumbel copulae to better represent tail dependence in the correlated samples. This vine copula performs better than the the Gaussian copula \cite{papaefthymiou2008using, xu2017power} since the latter has been proven to have $0$ tail dependence and, therefore, is less precise when describing the complicated dependence structures existing in renewable generation \cite{matthias2017simulating,becker2017generation}. 
Simulation  results carried out  on  the   IEEE  57-bus test system system reveal  that  our  proposed  method  can  accurately  estimate the probability density function (pdf) of the load margin  with  more than  two-order-of-magnitude  improvement in computing speed compared to the traditional MC method.


\vspace{-0.3cm}
\section{Problem Formulation}
This section formulates the probabilistic load-margin assessment problem. 

Let us first formulate the power system forward model as
\begin{equation}
    \label{forwardcpfmodel}
    y=f(\bm{x}). 
\end{equation}
Here, $y$ stands for the quantity of interest (QoI), which, in our case, is the load margin of a bus;  
$\bm{x}=[x_{1},x_{2},\ldots,x_{p}]$ is a vector of uncertain model parameters described by some distribution functions with finite variance. In our work, the active power and reactive power of the loads are considered to follow a Gaussian distribution and the wind power generation are assumed to follow the Weibull distribution; the ${f(\cdot)}$ is the nonlinear function that represents the  continuation power system model, which maps the model parameters, $\bm{x}$, to the QoI, $y$.  The detailed implementation step has been described by Ajjarapu \cite{ajjarapu2007computational}.

To obtain the probabilistic description of the load margin $y$ under these uncertain model parameters, a typical MC method draws a large number of  $N_\text{sample}$ samples, $\{ \bm{x}^{(j)} \}^{N_\text{sample}}_{j=1}$, that not only reflect the pdfs of the input parameters but also the correlation between them. Then, for each $\bm{x}^{(j)}$, $j=1,\dots,{N_\text{sample}}$, 
$
y^{(j)}=f(\bm{x}^{(j)}) 
$
is solved to get ${N_\text{sample}}$ load-margin solutions, $\{ y^{(j)} \}^{N_\text{sample}}_{j=1}$, from which the pdf of the load margin is determined. 
The CPF method is typically employed in power systems despite the fact that even a single evaluation of ${f(\cdot)}$ will involve multiple prediction and correction steps to obtain the load margin, $y$, and, hence, is admittedly a complicated, time-consuming solver---not to mention that $N_\text{sample}$ is typically required to be a large number in the MC sampling to ensure good computing accuracy.  Therefore, the goal of this paper is to greatly reduce the computational time of this method and to precisely model the correlation between the uncertain parameters as model inputs. 

\vspace{-0.45cm}
\section{Uncertainty Modeling}
In this section, we present the way to generate dependent high-dimensional samples as model inputs via vine copula. 

\subsubsection{Copula}
Recently, copulae have been proven to be successful in many industrial and financial applications for modeling the dependency between random inputs \cite{becker2017generation,konstantelos2018using}.
According to Sklar's theorem, any joint multivariate cumulative distribution function $F_\mathbf{X}$ of a $p$-dimensional random vector can be expressed in terms of its marginal distributions and a copula to represent their dependence. Formally, we have
\begin{equation}
    \label{copula}
    F_\mathbf{X}(\mathbf{x})=C(F_{X_{1}}(x_{1}), F_{X_{2}}(x_{2}), \dots, F_{X_{p}}(x_{p}) ). 
\end{equation}
Here, $F_{X_{i}}(x_{i})$ is the $i$th input marginal and $C(\cdot)$ is a copula that describes the dependence structure between the $p$-dimensional input variables \cite{nelsen2007introduction}. 
Accordingly, its joint multivariate density function, $f_\mathbf{X}$, can be obtained via
\begin{equation}
    \label{copuladensity}
    f_\mathbf{X}(\mathbf{x})=c(F_{X_{1}}(x_{1}), \dots, F_{X_{p}}(x_{p}) )\prod_{i=1}^{p} f_{i}(x_{i}). 
\end{equation}
Here, $c$ is the $p$-variate copula density and $f_{i}(x_{i})$ is the marginal density for $i$th variable. Since there exist different copula families, the choice of the copula function will influence the accuracy of the dependence modeling. 
The Gaussian copula is advantageous in certain applications thanks to its ability to generate high-dimensional correlated samples~\cite{xu2017power}.  Archimedean copulas are more useful in scenarios which require nonlinear tail dependence modeling. However, they are generally not scalable due to being limited to the bivariate case \cite{matthias2017simulating, wang2017probabilistic}. To overcome these shortcomings, we resort to vine copula next.

\subsubsection{Vine Copula}
Being a powerful tool in simulating high-dimensional correlated samples with various types of tail-dependence structures involved, vine copula is known for its capability of decomposing a multivariate density function into a cascade of bivariate pair copulae  \cite{joe1996families,matthias2017simulating}. Starting from the factorization on the joint density function, we get
\begin{equation}
\resizebox{\linewidth}{!}{
\begin{math}
\begin{aligned}
\label{factorization}
f_\mathbf{X}(\mathbf{x})=&f_{p}(x_{p})\cdot f_{{p-1}|{p}}(x_{p-1}|x_{p})\cdot
f_{{p-2}|{p-1},{p}}(x_{p-2}|x_{p-1},x_{p}) \\
&   \cdot \cdots f_{1|2,\dots,p}(x_{1}|x_{2},\dots, x_{p})\\
&=\prod_{i=1}^{p} f_{i|{i+1,\dots,p}}(x_{i}|x_{i+1}, \dots,x_{p} ). 
\end{aligned}
\end{math}}
\end{equation}
Based on the property that all the conditioned pdfs in \eqref{factorization} can be further transformed into the product of only bivariate copulae and one-dimensional density, e.g., $f_{2|1}(x_{2}|x_{1})=c_{2|1}(F_{2},F_{1})
\cdot f_{2}(x_{2})$, $f_{3|1,2}(x_{3}|x_{1},x_{2})=c_{3, 2|1}(F_{3|1},F_{2|1})
\cdot c_{3, 1}(F_{3},F_{1})\cdot
f_{3}(x_{3})$, and so on for the higher-dimensional cases \cite{matthias2017simulating}, it is easy to infer that the joint  density, $f_\mathbf{X}$, can be decomposed into a form that involves only bivariate copulae and marginal densities. Let us take $p=4$ as an example. Specifically, $f_\mathbf{X}(x_{1},x_{2},x_{3},x_{4})$ can be decomposed as
\begin{equation}
\resizebox{1.1\linewidth}{!}{
\begin{minipage}{1.35\linewidth}
\begin{math}
\begin{split}
\label{4factorization}
f_\mathbf{X}(x_{1},x_{2},x_{3},x_{4})=&f_{1}(x_{1})\cdot f_{2}(x_{2})\cdot f_{3}(x_{3})\cdot f_{4}(x_{4})
\cdot c_{1,2}(F_{1}(x_{1}),F_{2}(x_{2}))
\\
&
  \cdot c_{2,3}(F_{2}(x_{2}),F_{3}(x_{3}))
  \cdot c_{3,4}(F_{3}(x_{3}),F_{4}(x_{4}))
\\
&
\cdot c_{1,3|2}(F_{1|2}(x_{1}|x_{2}),F_{3|2}(x_{3}|x_{2}))
\\
&
\cdot c_{2,4|3}(F_{2|3}(x_{2}|x_{3}),F_{4|3}(x_{4}|x_{3}))
\\
&
\cdot c_{1, 4|2, 3}(F_{1|2, 3}(x_{1}|x_{2}, x_{3}),F_{4|2, 3}(x_{4}|x_{2},x_{3})). 
\end{split}
\end{math}
\end{minipage}
}
\end{equation}
The decomposition enables us to use multiple bivariate copulae to precisely describe the high-dimensional data structure.

However, it is worth pointing out that the order of pairwise conditioning on \eqref{factorization} and \eqref{4factorization} is not unique. Thus, we need a systematic way to decompose it and provide a unique solution. Two popular choices are the canonical vine (C-vine) and the drawable vine (D-vine). Both of them make use of a graphical tool to facilitate their decomposition into a set of cascade copula densities forming $p-1$ trees.  For the C-vine copula, a $p$-dimensional joint density is decomposed as
\begin{equation}
    \label{Cvine}
    f_\mathbf{X}(x_{1},\dots,x_{p})=\prod_{i=1}^{p} f_{i}(x_{i}) \prod_{j=1}^{p-1} \prod_{i=1}^{p-j} c_{j,j+i|1,\dots,j-1}.
\end{equation}
Similarly,  $f_\mathbf{X}$ is decomposed via the D-vine copula as 
\begin{equation}
    \label{Dvine}
  f_\mathbf{X}(x_{1},\dots,x_{p})=\prod_{i=1}^{p} f_{i}(x_{i}) \prod_{j=1}^{p-1} \prod_{i=1}^{p-j} c_{i,j+i|i+1,\dots,i+j-1}. 
\end{equation}
Here, $c_{j,j+i|1,\dots,j-1}$ is short for
$\scriptstyle c_{j,j+i|1,\dots,j-1}(F(x_{j}|x_{1},\dots,x_{j-1}),\allowbreak F(x_{i+j}|x_{1},\dots,x_{j-1}))$  and $c_{i,j+i|i+1,\dots,i+j-1}$ is short for \resizebox{\linewidth}{!}{$c_{i,j+i|i+1,\dots,i+j-1}(F(x_{i}|x_{i+1},\dots,x_{i+j-1}), F(x_{i+j}|x_{i+1,\dots,x_{i+j-1}}))$}. 
A simple graph demo for a $4$-dimensional C-vine and D-vine is displayed in Fig. 1. For more details about their descriptions and implementations, the reader is referred to \cite{matthias2017simulating}.  It is also easy to see that \eqref{4factorization} is  obtained from the D-vine copula.  Using this vine-copula technique, we are able to generate the correlated samples that reflect the precise dependent structures of the model inputs such as loads and renewables.  
\begin{figure}[!htbp]
\vspace{-0.2cm}
\centering
\includegraphics[width=0.48\textwidth]{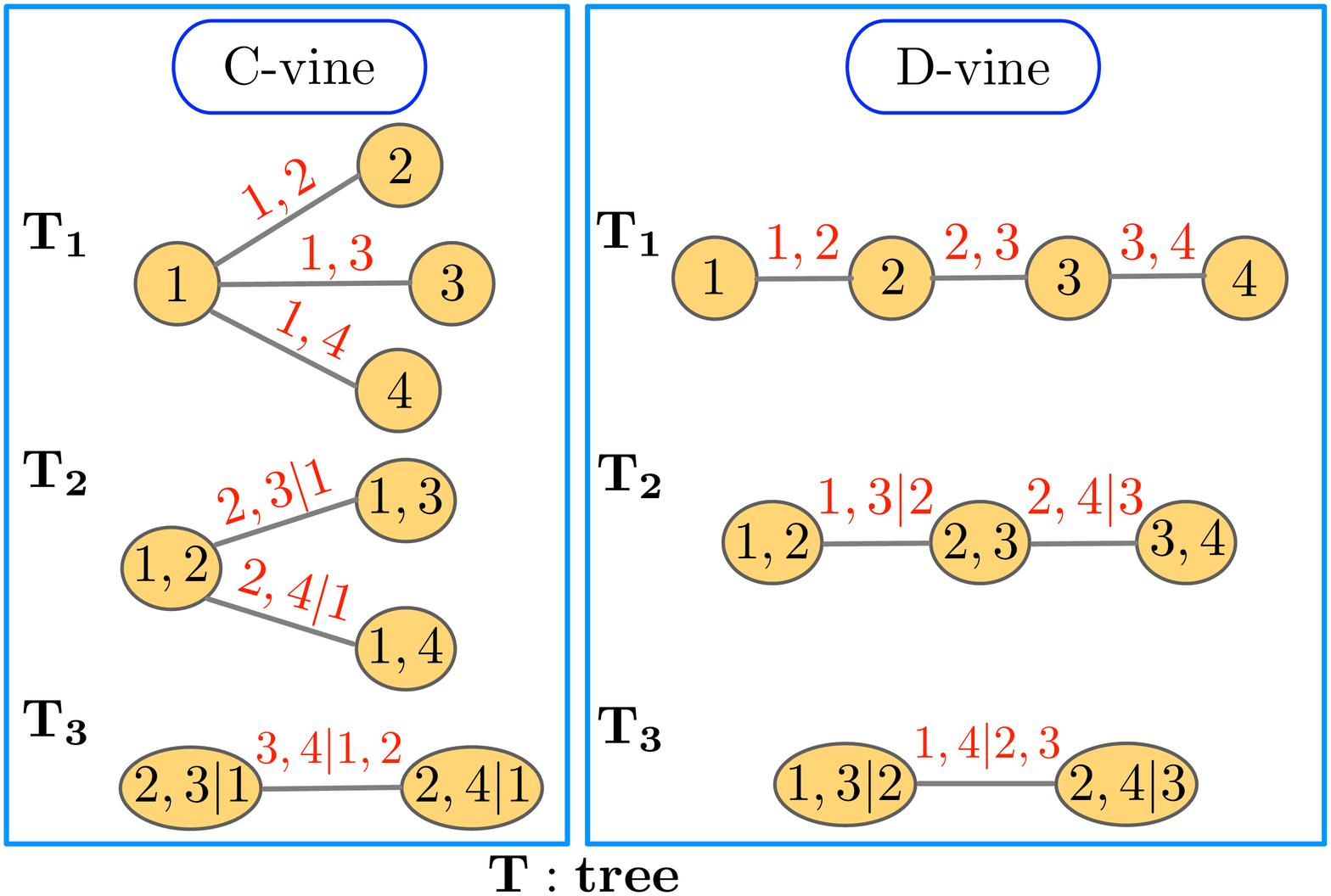}
 \setlength{\abovecaptionskip}{-5pt}
 \caption{Examples for a $4$-dimensional C-vine and D-vine with pair copulae (\textcolor{red}{red}).}
 \label{fig:1}
\vspace{-0.3cm} 
 \end{figure}

\vspace{-0.2cm}
\section{Reduced-Order Modeling}
In this section, we present a nonparametric, reduced-order modeling technique using GPE.
\vspace{-0.3cm}
\subsection{Problem Description}
Let us first formulate the probabilistic load-margin assessment problem in the GPE framework. Here, the CPF model is denoted by ${f(\cdot)}$. Its corresponding vector-valued random input of $p$ dimensions is denoted as $\mathbf{x}$, which accounts for the uncertainties from the variations of the loads and the renewable energy generation. Due to the randomness of $\mathbf{x}$, we may observe $n$ samples as a  finite collection of the model input as $\{\mathbf{x}_{1}, \mathbf{x}_{2}, \dots, \mathbf{x}_{n} \}$. 
Accordingly, its evaluated model output $f(\mathbf{x})$, i.e., load margin, also becomes random, and has its corresponding $n$ realizations denoted by $\{f(\mathbf{x}_{1}), f(\mathbf{x}_{2}), \dots, f(\mathbf{x}_{n})\}$. If we assume that the model output is a realization of a Gaussian process, then the finite collection, $\{f(\mathbf{x}_{1}), f(\mathbf{x}_{2}), \dots, f(\mathbf{x}_{n})\}$, of the random variables, $f(\mathbf{x})$, will follow a joint multivariate normal probability distribution as
\begin{equation}
\label{eqn:2}
   \left[\begin{array}{c}{{\scriptstyle f\left(\mathbf{x}_{1}\right)}} \\ {\vdots} \\ {{\scriptstyle f\left(\mathbf{x}_{n}\right)}}\end{array}\right] \sim \scriptstyle{ \mathcal{N}} \left(\left[\begin{array}{c}{{\scriptstyle m\left(\mathbf{x}_{1}\right)}} \\ {\vdots} \\ {{\scriptstyle m\left(\mathbf{x}_{n}\right)}}\end{array}\right],\left[\begin{array}{ccc} {{\scriptstyle k\left(\mathbf{x}_{1}, \mathbf{x}_{1}\right)}} & {\cdots} & {{\scriptstyle k\left(\mathbf{x}_{1}, \mathbf{x}_{n}\right)}} \\ {\vdots} & {\ddots} & {\vdots} \\ {{\scriptstyle k\left(\mathbf{x}_{n}, \mathbf{x}_{1}\right)}} & {\cdots} & {{\scriptstyle k\left(\mathbf{x}_{n}, \mathbf{x}_{n}\right)}}\end{array}\right]\right).
\end{equation}
Here, let us denote $\bm{m}(\bm{\cdot})$ as the mean function and $\bm{k}(\bm{\cdot},\bm{\cdot})$ as a kernel function that represents the covariance function. Then, \eqref{eqn:2} can be simplified as
\begin{equation}
\label{eqn:3}
\bm{f}\left(\mathbf{X}\right) | \mathbf{X} \sim \mathcal { N }\left(\bm{m}\left(\mathbf{X} \right), \bm{k}\left(\mathbf{X}, \mathbf{X}\right)\right),
\end{equation}
where $\mathbf{X}$ is an $n\times p$ matrix, denoted by $[\mathbf{x}_{1}, \mathbf{x}_{2}, \dots, \mathbf{x}_{n}]^{\transpose}$;
$\bm{f}(\mathbf{X})$ stands for $[f(\mathbf{x}_{1}), f(\mathbf{x}_{2}), \dots, f(\mathbf{x}_{n})]^{\transpose}$; and $\bm{m}(\mathbf{X})$ represents $[m(\mathbf{x}_{1}), m(\mathbf{x}_{2}), \dots, m(\mathbf{x}_{n})]^{\transpose}$.

Now, if an independent and identically distributed (i.i.d.) Gaussian noise $\boldsymbol{\varepsilon} \sim \mathcal{N}(0, \sigma^{2}\mathbf{I}_{n})$
 (where $\mathbf{I}_{n}$ and $\sigma^{2}$ are an $n$-dimensional identity matrix and the variance, respectively) is considered in the system output, 
 $\bm{f}(\mathbf{X})$,  the observations $\mathbf{Y}$ will be expressed as
 
\begin{equation}
\label{eqn:5}
\mathbf{Y} | \mathbf{X} \sim \mathcal { N }\left(\bm{m}\left(\mathbf{X}\right), \bm{k}\left(\mathbf{X}, \mathbf{X}\right) + \sigma^{2}\mathbf{I}_{n} \right).
\end{equation}
Note that $\boldsymbol{\varepsilon}$ is also called a ``nugget''. If $\sigma^{2} = 0$, then  $f(x)$ is observed without noise.  However, in practical implementation, the nugget is always added for the sake of numerical stability.

\vspace{-0.45cm}
\subsection{Bayesian Inference}
 Here, we present the way to use the abovementioned finite collection of $n$ samples, $(\mathbf{Y}, \mathbf{X})$, to infer the unknown system output,  $\mathbf{y}(\mathbf{x})$, on the sample space of $\mathbf{x} \in \mathbb{R}^{p}$ in a Bayesian-inference framework.  

It is well-known that a Bayesian posterior distribution of the unknown system output can be inferred from a Bayesian prior distribution of $\mathbf{y}(\mathbf{x})$ and the likelihoods obtained from the observations. Let us first assume a Bayesian prior distribution of $\mathbf{\mathbf{y}(\mathbf{x})} | \mathbf{x} $, expressed as
\begin{equation}
\label{eqn:6}
\mathbf{\mathbf{y}(\mathbf{x})} | \mathbf{x} \sim \mathcal { N }\left(\bm{m}\left(\mathbf{x}\right), \bm{k}\left(\mathbf{x}, \mathbf{x}\right) + \sigma^{2}\mathbf{I}_{n_{x}} \right). 
\end{equation}
Combined with the observations provided by the finite collection of the samples $\{\mathbf{Y}, \mathbf{X}\}$, we can formulate the joint distribution of $\mathbf{Y}$ and $\mathbf{\mathbf{y}(\mathbf{x})} | \mathbf{x}$ as
\begin{equation}
\label{eqn:7}
\left[\begin{array}{c}{{\mathbf{Y}}}  \\ \mathbf{\mathbf{y}(\mathbf{x})} | \mathbf{x} \end{array}\right] \sim { \mathcal{N} \left(\left[\begin{array}{c}{\bm{m}\left(\mathbf{X}\right)}  \\  \bm{m}\left(\mathbf{x}\right)\end{array}\right],\left[\begin{array}{cc} { \mathbf{K}_{11} }  & {\mathbf{K}_{12}} \\   {\mathbf{K}_{21}} & {\mathbf{K}_{22}} \end{array}\right]\right),
}
\end{equation}
where $ \mathbf{K}_{11} = \bm{k}\left(\mathbf{X}, \mathbf{X}\right) + \sigma^{2}\mathbf{I}_{n}$; $\mathbf{K}_{12} = \bm{k}\left(\mathbf{X}, \mathbf{x}\right)$; $\mathbf{K}_{21} = \bm{k}\left(\mathbf{x}, \mathbf{X}\right)$; and $\mathbf{K}_{22} = \bm{k}\left(\mathbf{x}, \mathbf{x}\right) + \sigma^{2}\mathbf{I}_{n_{x}}$. 

Now, using  the  rules of the conditioned Gaussian distribution  \cite{eaton1983multivariate}, we can infer the Bayesian posterior distribution of the system output  $\mathbf{y}(\mathbf{x})$ conditioned upon the observations $\left(\mathbf{Y}, \mathbf{X}\right)$. It follows a Gaussian distribution given by
\begin{equation}
\label{eqn:8}
\mathbf{y}(\mathbf{x}) | \mathbf{x}, \mathbf{Y}, \mathbf{X} \sim \mathcal { N }\left( \bm{\mu} \left(\mathbf{x} \right), \bm{\Sigma} \left( \mathbf{x} \right) \right),
\end{equation}
where 
\begin{equation}
\label{gpemean}
    \bm{\mu} \left(\mathbf{x} \right) = \bm{m}(\mathbf{x}) + \mathbf{K}_{21}\mathbf{K}_{11}^{-1} (\mathbf{Y} - \bm{m}(\mathbf{X})), 
\end{equation}
\begin{equation}
\label{gpecov}
    \bm{\Sigma} \left( \mathbf{x} \right) =  \mathbf{K}_{22} - \mathbf{K}_{21} \mathbf{K}_{11}^{-1} \mathbf{K}_{12}.
\end{equation}
To this point, the general form of the GPE has been derived. On one hand, we can directly use  \eqref{gpemean} as a surrogate model (a.k.a. the response surface or reduced-order model) to very closely capture the behavior of the nonlinear CPF model while being computationally inexpensive to evaluate. On the other hand, we may use \eqref{gpecov} to quantify the uncertainty of the surrogate itself. In this paper, we only need to use \eqref{gpemean} as a surrogate model.
 
 

\vspace{-0.45cm}
\subsection{Mean and Covariance Functions}
Let us describe the mean function $\bm{m}(\bm{\cdot})$
and the covariance function represented via the kernel $\bm{k}(\bm{\cdot},\bm{\cdot})$ that characterizes the GPE.
The mean function models the prior belief about the existence of a systematic trend expressed as
\begin{equation}
\label{trend}
    \bm{m}(\mathbf{x}, \boldsymbol{\beta}) = \mathbf{H}(\mathbf{x}) \boldsymbol{\beta}.
\end{equation}
Here, $\mathbf{H}(\mathbf{x})$ can be any set of basis functions. For example, let $ \mathbf{x}_{i} = [x_{i1},\dots,x_{ip}]$ be the $i$th sample, where $i=1,2,\dots,n$, wherein $x_{ik}$ represents its $k$th element, where $k=1,2,\dots, p$. For instance, $\mathbf{H}(\mathbf{x}_{i}) = 1$ is a constant basis; $\mathbf{H}(\mathbf{x}_{i}) =[1, x_{i 1},\dots, x_{i p}]$ is a linear basis; $\mathbf{H}(\mathbf{x}_{i}) = [1, x_{i 1}, \dots, x_{i p}, x_{i 1}^{2},\dots, x_{i p}^{2}]$ is a pure quadratic basis, and $\boldsymbol{\beta}$ is a vector of hyperparameters. 

Since the covariance function is represented by a kernel function, choosing the latter is a must. Popular choices include the square exponential kernel ($k_{\text{SE}}$), the exponential kernel ($k_{\text{E}}$), the rational quadratic kernel ($k_{\text{RQ}}$), and the Martin 3/2 kernel ($k_{3/2}$); they are listed in Table~\ref{t: 6}. As for the parameters of a kernel function, they are defined as follows: $\tau$ and $\ell_{k}$ are the hyperparameters defined in the positive real line; $\sigma^{2}$ and $\ell_{k}$ correspond to the order of the magnitude and the speed of variation in the $k$th input dimension, respectively. Let $\boldsymbol{\theta} = [\tau, \ell_{1}, \dots, \ell_{p}]$ contain the hyperparameters of the covariance function, i.e.,
\begin{equation}
\label{kerneltype}
k\left(\mathbf{x}_{i}, \mathbf{x}_{j} | \boldsymbol{\theta} \right)  = \Cov(\mathbf{x}_{i}, \mathbf{x}_{j} | \boldsymbol{\theta}).
\end{equation}

Until now, the model structure of the GPE has been fully defined. For simplicity, we write $\boldsymbol{\eta} = (\sigma^{2}, \boldsymbol{\beta}, \boldsymbol{\theta})$ to represent all the hyperparameters in the GPE model.

\begin{table}[th]
\vspace{-0.5cm}
\renewcommand{\arraystretch}{1.3}
\caption{ Commonly Used Covariance Kernels for Gaussian Process}
\label{t: 6}
\vspace{-0.2cm}
\centering
\begin{tabular}{rll}
\toprule
\hline

$k_{\text{SE}}\left(\mathbf{x}_{i}, \mathbf{x}_{j}\right)$ & $\tau^{2} \exp \left(- \sum\limits_{k=1}^{p} \frac{r_{k}^{2}}{2 \ell_{k}^{2}}\right) $ \\
$k_{\text{E}}\left(\mathbf{x}_{i}, \mathbf{x}_{j}\right)$ & $\tau^{2} \exp \left(- \sum\limits_{k=1}^{p} \frac{\left| r_{k} \right|}{ \ell_{k}}\right)$ \\ 
$k_{\text{RQ}}\left(\mathbf{x}_{i}, \mathbf{x}_{j}\right)$& $\tau^{2}\left(1+ \sum\limits_{k=1}^{p} \frac{r_{k}^{2}}{2 \alpha \ell_{k}^{2}}\right)^{ - \alpha}$ \\ 
$k_{3/2}\left(\mathbf{x}_{i}, \mathbf{x}_{j} \right)$& $\tau^{2}\left(1+ \sum\limits_{k=1}^{p} \frac{\sqrt{3} r_{k}}{\ell_{k}}\right) \exp \left(-\sum\limits_{k=1}^{p} \frac{\sqrt{3} r_{k}}{\ell_{k}}\right)$ \\ 
&  $(r_{k} = \left|x_{ik}-x_{jk}\right|)$ \\
\bottomrule
\end{tabular}
\vspace{-0.25cm}
\end{table}

\vspace{-0.15cm}
\section{Probabilistic Load-Margin Assessment}
\vspace{-0.15cm}
Here, we illustrate the steps for conducting the probabilistic load-margin assessment using the GPE.

\subsubsection{Training Sample Generation}
In order to acquire the GPE-based surrogate described in (\ref{gpemean}), we need to obtain the observation sets contained in $\left(\mathbf{Y}, \mathbf{X}\right)$. To obtain the system realization $\mathbf{Y}$, we must generate $n$ samples, $\mathbf{X}$, that will be evaluated through the CPF model, ${f(\cdot)}$. To avoid long training time of the GPE, $n$ should be small. To meet this requirement, the Latin hypercube sampling is typically chosen. It generates near-random samples and, therefore, has a faster convergence rate than the MC sampling, which generates pure random samples. This is especially true in our case since $n$ needs to be small~\cite{santner2003design} and, therefore, the MC sampling  based on the central limit theorem is not suggested here. Note that the Latin hypercube sampling generates i.i.d. samples while the correlation between renewable energy generations are inevitable. Therefore, we need to use the aforementioned vine copula to further transform these i.i.d. samples into the correlated ones to improve the training performances.

\subsubsection{GPE Construction} With $(\mathbf{Y}, \mathbf{X})$, we can estimate the hyperparameters $\bm{\eta}$ in the GPE. Following Gelman \textit{et al.}  \cite{gelman2014bayesian}, we choose to adopt the Gaussian maximum likelihood estimator (MLE) since it is the most efficient estimator under a Gaussian distribution, which is followed by the calculated residuals, and it is easy to compute.  First, to indicate the hyperparameters, let us rewrite \eqref{eqn:5} as
\begin{equation}
\label{reform}
\mathbf{Y} | \mathbf{X},\boldsymbol{\eta} \sim \mathcal { N }\left(\bm{m}\left(\mathbf{X}\right), \bm{k}\left(\mathbf{X}, \mathbf{X}\right) + \sigma^{2}\mathbf{I}_{n} \right). 
\end{equation}
Then, using the Gaussian MLE, we obtain
\begin{equation}
\label{eqn:11}
    \widehat{\boldsymbol{\eta}} = \left(\widehat{\boldsymbol{\beta}}, \widehat{\boldsymbol{\theta}}, \widehat{\sigma}^{2} \right)=\underset{\boldsymbol{\beta}, \boldsymbol{\theta}, \sigma^{2}}{\arg \max } \log P\left(\mathbf{Y} | \mathbf{X}, \boldsymbol{\beta}, \boldsymbol{\theta}, \sigma^{2}\right).
\end{equation}
Using \eqref{trend}--\eqref{reform} and using $\mathbf{H}$ instead of  $\mathbf{H}(\mathbf{x})$ for simplicity, the marginal log-likelihood can be expressed as
\begin{equation}
\label{eqn:12}
\begin{aligned} & \log P\left(\mathbf{Y} | \mathbf{X}, \boldsymbol{\beta}, \boldsymbol{\theta}, \sigma^{2} \right) \\
=&-\frac{1}{2}(\mathbf{Y}-\mathbf{H} \boldsymbol{\beta})^{\transpose}\left[\bm{k}(\mathbf{X}, \mathbf{X} | \boldsymbol{\theta})+\sigma^{2} \mathbf{I}_{n}\right]^{-1}(\mathbf{Y}-\mathbf{H} \boldsymbol{\beta}) \\ &-\frac{n}{2} \log 2 \pi-\frac{1}{2} \log \left|\bm{k}(\mathbf{X}, \mathbf{X} | \boldsymbol{\theta})+\sigma^{2} \mathbf{I}_{n}\right|, 
\end{aligned}
\end{equation}
which implies that the Gaussian MLE of $\boldsymbol{\beta}$ conditioned on $\boldsymbol{\theta}$ and $\sigma^{2}$ is a weighted least-squares estimator given by 
\begin{equation}
\label{eqn:13}
\resizebox{\linewidth}{!}{ $\hat{\boldsymbol{\beta}}\left(\boldsymbol{\theta}, \sigma^{2}\right)= \left[\mathbf{H}^{\transpose}\left[\bm{k}(\mathbf{X}, \mathbf{X} | \boldsymbol{\theta})+\sigma^{2} \mathbf{I}_{n}\right]^{-1} \mathbf{H}\right]^{-1} \mathbf{H}^{\transpose}\left[\bm{k}(\mathbf{X}, \mathbf{X} | \boldsymbol{\theta})+\sigma^{2} \mathbf{I}_{n}\right]^{-1} \mathbf{Y}$}.
\end{equation}
Since $\widehat{\boldsymbol{\beta}} $ is a function of $\left(\widehat{\boldsymbol{\theta}}, \widehat{\sigma}^{2}\right)$,
let us insert \eqref{eqn:13} into \eqref{eqn:12} to reduce the number of the hyperparameters. Then, \eqref{eqn:11} is further simplified as
\begin{equation}
\label{eqn:14}
    \left(\widehat{\boldsymbol{\theta}}, \widehat{\sigma}^{2}\right) =\underset{\boldsymbol{\theta}, \sigma^{2}}{\arg \max } \log P\left(\mathbf{Y} | \mathbf{X}, \hat{\boldsymbol{\beta}}\left(\boldsymbol{\theta}, \sigma^{2}\right), \boldsymbol{\theta}, \sigma^{2}\right).  
\end{equation}
Now, we only need to estimate the hyperparameters $\left(\widehat{\boldsymbol{\theta}}, \widehat{\sigma}^{2}\right)$ instead of $\left(\widehat{\boldsymbol{\beta}}, \widehat{\boldsymbol{\theta}}, \widehat{\sigma}^{2} \right)$.   Then, we utilize  a gradient-based optimizer to solve this optimization as described in \cite{rasmussen2006gaussian}. 
Once $\widehat{\boldsymbol{\eta}}$ is obtained, the GPE model is fully constructed.  More details can be found in \cite{gelman2014bayesian}. 
\subsubsection{Sample Evaluation} 
Now, we can execute an MC sampling procedure to generate a large amount of samples and transform them into the correlated ones through a vine copula. These large amount of samples can be evaluated through the GPE-based surrogate expressed in (\ref{gpemean}) at almost no computational cost. Finally, the pdf of the system response can be obtained.

\vspace{-.4cm}
\section{Simulation Results}
A case study is conducted using the IEEE 57-bus system\cite{PSTCA}. 
The algorithms are tested with \textsc{Matpower} package using MATLAB\textsuperscript{\textregistered} R$2018$a version on a  desktop with $3.50$-GHz Intel\textsuperscript{\textregistered} Xeon(R) CPU E5-1650 v2 processors and a $32$ GB of main memory.

The uncertain inputs include the variations in the loads as well as in the wind  generation. Here, it is assumed that the loads follow a Gaussian distribution with mean values equal to the original bus loads and standard deviations equal to $5\%$ of their means. Four wind farms are connected to Buses 16, 17, 47, and 48 whose generation profiles follow the Weibull distribution with the shape and scale parameter set as $\{2.06,7.41\}$, $\{2.1,7.2\}$, $\{2.06,7.41\}$, and $\{2.3,7.2\}$, respectively. To reflect the correlation between these five uncertain inputs, we use the D-vine copula to generate these samples since Becker claims that the latter provides the highest accuracy when simulating wind power generations\cite{becker2017generation}. Furthermore, for our problem with five-dimensional inputs, we will have $p(p-1)/2=10$ pair-copulae. Although their structure can be obtained from the D-vine, we still need to estimate the copula family for every pair copula (e.g., Frank, Gumbel), and the parameters in every copula. The copula family and copula parameters can be estimated from real data sets using a maximum likelihood estimator as described in 
\cite{matthias2017simulating}. Here, for simplicity, we assume we know the copula family and the copula parameter for each pair-copula. As suggested by Becker \cite{becker2017generation}, the predominant copula family is the Frank copula for wind-data dependence simulation.  Therefore, we assume that we have $7$ Frank copulae, $2$ Gaussian copulae, and $1$ Gumbel copula. The simulated samples for the four wind farms are displayed in Fig. 2. It shows that both the upper-tail and lower-tail dependence can be simulated using the D-vine copula. 

\begin{figure}[!htbp]
\vspace{-0.5cm}
\centering
\includegraphics[scale=0.38]{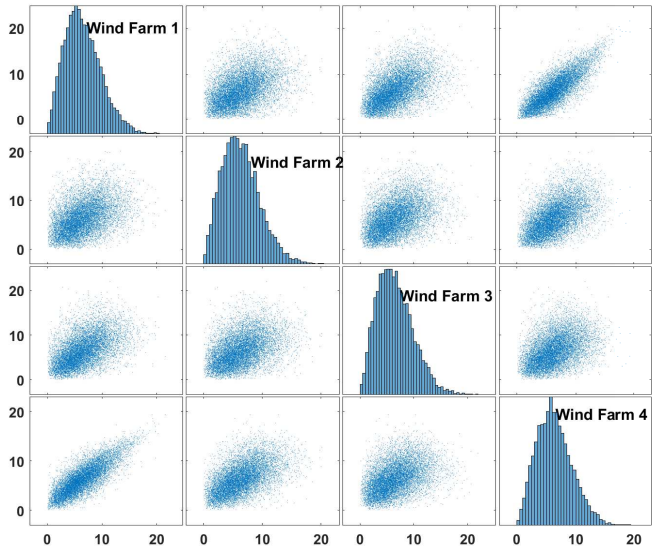}
 \setlength{\abovecaptionskip}{-25pt}
 \vspace{-.5cm}
 \caption{$1$-d histograms and scatter plots for wind farm samples (in MVA).}
 \label{fig:1}
\vspace{-0.25cm} 
 \end{figure}


Let us choose the load margin at Bus 25 as the target quantity of interest. The simulation results obtained with the MC and the GPE methods are provided in Fig. 3. The simulation results obtained with the MC method with $10,000$ samples are used as a benchmark to validate the GPE-based method. It can be seen that with only $15$ training samples, the GPE method with a pure quadratic basis functions can provide highly accurate simulation results under different kernel functions, such as $k_{\text{SE}}$ and $k_{3/2}$. The proposed GPE-based method significantly reduces the computation time over the traditional MC method without any loss of accuracy.

\begin{figure}[!htbp]
\vspace{-0.3cm}
\centering
\includegraphics[width=.48\textwidth]{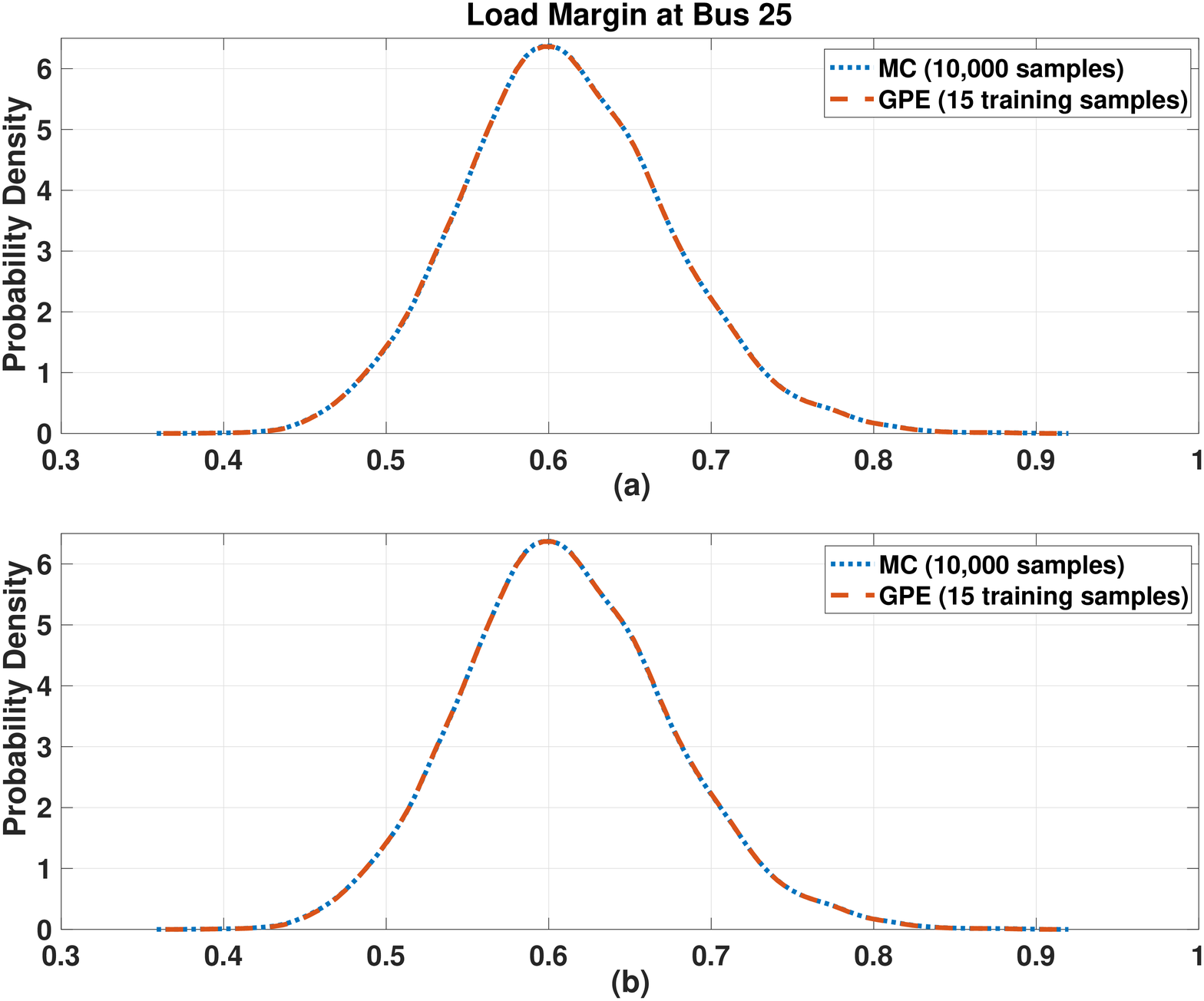}
 \setlength{\abovecaptionskip}{-5pt}
 \caption{Probability density plots for load margin at Bus 25 using GPE (a)  with a pure quadratic basis and $k_{3/2}$ kernel, and (b) with a pure quadratic basis and $k_{\text{SE}}$ kernel.}
 \label{fig:1_old}
\vspace{-0.25cm} 
 \end{figure}

\vspace{-0.2cm}
\section{Conclusions and Future Work}
\vspace{-0.15cm}
In this paper, we propose a novel GPE-based method for the PPF analysis.  The GPE serves as a reduced-order model for the nonlinear CPF that enables an evaluation of the time-consuming CPF solver at the sampled values with a negligible computational cost. The simulation results reveal that the proposed method exhibits an impressive performance as compared to the traditional MC method. As a future work, we will use real data to estimate the copula family and copula parameters of the vine copula and will attempt to improve the performance of the proposed method when it is applied to very-large-scale power systems.
\vspace{-0.2cm}

\bibliographystyle{IEEEtran}
\bibliography{IEEEabrv,References}

\end{document}